\documentclass[a4paper,10pt]{article}

\usepackage{amssymb}
\usepackage{amsmath}
\usepackage[usenames,dvipsnames]{color}
\usepackage{hyperref}
\usepackage[left=1in,right=1in,top=1in,bottom=1in]{geometry}       

\title{The Coupling of Shape Dynamics to Matter}
\author{\bf Henrique Gomes\footnote{\href{mailto:gomes.ha@gmail.com}{gomes.ha@gmail.com}}\\\it Department of Physics,  University of California, Davis,   CA, 95616}

\let\oldmarginpar\marginpar
\renewcommand\marginpar[1]{\oldmarginpar{\color{red}\raggedright\scriptsize #1}}

\newcommand{\mean}[1]{\ensuremath{\lf\langle #1 \rt\rangle }}

\def\be{\begin{equation}}
\def\ee{\end{equation}}
\def\bea{\begin{eqnarray}}
\def\eea{\end{eqnarray}}

\def\lf {\ensuremath{\left}}
\def\rt {\ensuremath{\right}}

\begin{document}

\maketitle

\begin{abstract}
 Shape Dynamics (SD) is a theory dynamically equivalent to vacuum General Relativity (GR), which has a different set of symmetries. It trades refoliation invariance, present in GR, for local 3-dimensional conformal invariance. This contribution to the Loops 11 conference addresses one of the more urgent questions regarding the equivalence: is it possible to incorporate normal matter in the new framework? The answer is yes, in certain regimes. We present general criteria for coupling and apply it to a few examples.The outcome presents bounds and conditions on scalar densities (such as the Higgs potential and the cosmological constant) not present in GR.
\end{abstract}

\section{Introduction}

 Shape Dynamics (SD) \cite{Gomes:2011zi,Gomes:2010fh} is a classical theory equivalent to pure gravity in which the spacetime picture that underlies GR is replaced by an intrinsically spatial picture. In particular, invariance under spacetime refoliations is traded for invariance of the theory  under local spatial conformal transformations that preserve the total spatial volume. The equivalence is obtained by manipulating the constraint structure of General relativity, but can only be achieved if one first uses the Stuckelberg mechanism to extend the original  phase space.  Much like other known uses of extended phase space\footnote{ We just mention two examples: the original Stueckelberg trick \cite{Stueckelberg}, which   introduces a scalar field into an Abelian gauge theory making it massive but  preserving gauge invariance,  and the BRST construction, which adds ghosts and ghost momenta to the original phase space in order to apply cohomology tools to understand the space of gauge invariant functionals.} the use of extended phase space for obtaining SD seems unavoidable. The basic conditions required for the symmetry trading is the existence of two (the original and the ``new") symmetries which are maximally symplectic (i.e. gauge fix one another). Motivations for SD are many, and discussed in  \cite{Gomes:2011zi,Gomes:2010fh}, but a ``uniqueness" type argument comes about only when considering field quantization. In fact, even before field quantization, in the absence of a true Hamiltonian (as in generally covariant systems), the fact that we have two symmetries gauge-fixing one another automatically implies that both symmetries are kept in the BRST formalism. Work on the construction of SD through ``symmetry doubling" will be reported on shortly. 

One of the upsides of the symmetry trading is that the constraint structure of SD becomes much simpler than that of GR, and in fact qualifies as a (infinite-dimensional) Lie algebra, as opposed to the Lie algebroid structure of GR. Shape dynamics is furthermore  left with a unique (and single) Hamiltonian constraint to generate global time translation, yielding a tool to attack  those problems of time related to refoliation invariance. However the trade-off implies that this global Hamiltonian is non-local. This holistic aspect of the theory  makes it more difficult to  incorporate matter as an ``add on" to the theory. Instead,  one must input the matter Hamiltonian  already at the base level of the construction of SD, in the linking theory. 

The problem of incorporating different interactions to the theory are of utmost importance, as it is through matter that one may obtain most phenomenological predictions  from the theory, particularly those that might differ from its general relativistic counterparts.    
It turns out that the construction of a SD extension with standard matter content can be obtained from the appropriate GR theory by straightforward application of the construction principle given in \cite{Gomes:2011zi}.
This paper is based on \cite{Gomes:2011au}. 

\section {Preliminaries}\label{sec:preliminaries}

To make this paper sufficiently self-contained, we briefly review the construction principle for equivalent gauge theories, in particular the construction of Shape Dynamics in vacuum. The following procedure is a summary of that contained in \cite{Gomes:2011zi}. Basically, the procedure in the concrete case of General Relativity consists in first  extending phase space to accommodate extra redundant symmetries obtained through ``conformalization". This does \emph{not} mean that the extended theory is in any way conformally invariant.  Then we apply two different particular gauge fixings and follow the Dirac analysis to obtain two dual theories with distinct symmetry properties: GR and Shape Dynamics.

\subsection{Pure Shape Dynamics}

Let us now briefly review the construction of Shape Dynamics as a theory equivalent to ADM gravity on a compact Cauchy surface $\Sigma$ without boundary. For a more thorough review see   \cite{Gomes:2011zi} or \cite{deA.Gomes:2011wk}. We start with the standard ADM phase space $\Gamma_{ADM}=\{(g,\pi):g\in \mathrm{Riem},\pi\in T_g^*(\mathrm{Riem})\}$, where $\mathrm{Riem}$ denotes the set of Riemannian metrics on the above defined 3-manifold $\Sigma$, and the usual first class ADM constraints, i.e. the scalar constraints $S(x)=\frac{\pi^{ab}\pi_{ab}-\frac{1}{2}\pi^2}{\sqrt g}-\sqrt g R$ and momentum constraints $H^a(x)=\pi^{ab}_{~;b}(x)$ thereon. We extend the ADM phase space with the phase space of a scalar field $\phi(x)$ and its canonically conjugate momentum density $\pi_\phi(x)$, which we introduce as additional first class constraints $\mathcal{Q}(x)=\pi_\phi(x)\approx 0$. The system is thus merely a trivial embedding of the original ADM onto the extended phase space. Now  use the canonical transformation $T_\phi$ generated by the generating functional $F=\int d^3x\left(g_{ab}e^{4\hat \phi}\Pi^{ab}+\phi\Pi_\phi\right)$,  where $\hat \phi(x):=\phi(x)-\frac 1 6 \ln\langle e^{6\phi}\rangle_g$ using the mean $\langle f\rangle_g:=\frac 1 V \int d^3x\sqrt{|g|} f(x)$ and 3-volume $V_g:=\int d^3x\sqrt{|g|}$. These ``means" and volumes technically appear as a manner to yield dynamically equivalent theories. They force the appearance of a global Hamiltonian in SD.

 At this point we have the first class set of constraints
$   T_\phi S(x),~
   T_\phi H(x)$ and $
   T_\phi Q(x)=\pi_\phi(x)-4(\pi-\langle\pi\rangle\sqrt g)(x)$. 
We then perform the phase space reduction for the gauge fixing condition $\pi_\phi(x)=0$. 
The crucial technical steps come once one has imposed such a gauge-fixing, also sometimes called ``best-matching constraint". For it is necessary for the symmetry trading to work that this gauge-fixing be completely second class wrt the set of original constraints one wants to trade for, and first class wrt all the other constraints. Furthermore one must show that the variable conjugate to $\pi_\phi$, i.e. $\phi$ can be set such that that same original set of constraints can be strongly solved for (solved for in the entire phase space). 

After some algebra, we find  that all but one of the $  T_\phi S(x)$ constraints can be solved for $\phi=\phi_o$, and this one leftover constraint yields the global Hamiltonian of SD. The end result is the following set of first class constraints, which define Shape Dynamics
\begin{equation}\label{equ:pureSDconstraints}
 \begin{array}{rcl}
   H_{SD}&=&T_{\phi_o} S(N_o)\\
   H(\xi)&=&\int d^3x \pi^{ab}\mathcal{L}_\xi g_{ab}\\
   Q(\rho)&=&\int d^3x \rho\left(\pi-\langle \pi\rangle\sqrt{|g|}\right),
 \end{array}
\end{equation}
where $\phi_o(g,\pi)$ is such that $H_{SD}=0$ combined with $\phi=\phi_o(g,\pi)$ is equivalent to $T_\phi S(x)=0$ (at the surface $\pi_\phi=0$). Moreover, $N_o$ denotes the CMC lapse function, $\pi=\pi^{ab}g_{ab}$ and $\langle\pi\rangle=\frac 1 V\int d^3 x\pi$. The constraints $H(\xi)$ generate spatial diffeomorphisms and the constraints $Q(\rho)$ generate conformal transformations that leave the total spatial volume invariant.

 The main issue that forbids or allows the trading is that the gauge-fixed constraints arising from the Stueckelberg mechanism (best-matching constraints)  are first-class amongst themselves (and possibly also with respect to other constraints that one does not want to disturb, e.g. 3-diffeomorphisms) and second class solely with respect to the constraint that one wants to trade for.  
Thus the construction of a linking theory is non-generic. The condition that (a subset of) the best matched original constraints can be uniquely solved for the auxiliary variables $\phi$ is indispensable for the existence of a linking theory. 

\section{General Procedure}\label{sec:generalProcedure}

\subsection{What constrains the coupling to the conformal factor?}
When we try to couple different fields to gravity, we will have to face basically one question: how does one scale fields? What exponent is viable in the conformal coupling of a scalar field,    $\psi\rightarrow e^{\alpha\hat\phi}\psi$? This is an important issue because if the scaling is not correct we could encounter two difficult obstructions.

 The first obstruction is that if we are dealing with a field that possesses some kind of gauge symmetry it might not be possible to find a constraint $\mathcal{Q}$  that is first class with respect to the gauge constraint. The second is that the conserved charge (which we call $D$) implicit inside $\mathcal{Q}$  that defines the foliation  might depend on the field. As a subset case, there might be an even worse consequence if the field in question possesses some gauge symmetry (like electromagnetism). For then the charge could turn out to depend on the gauge potential. This is possible in spite of the existence of a  field-dependent $D$ that is first class with respect to the gauge generator of the field. Indeed with any other choice of scaling than the one chosen in the text, this possibility is realized with any Yang-Mills type interaction for non ``neutral coupling".  
 Neutral coupling is the particular choice of \emph{empty} scaling $\alpha=0$, i.e. non-gravitational fields have conformal weight zero. This means that fields are only scaled in the sense that they are ``carried along" by the scaling of the spatial metric.

 \subsection{Explicit coupling}

We follow the construction principle for pure Shape Dynamics and suppose we have a gravity-matter system on a compact manifold without boundary,  with constraints
\begin{equation}
 \begin{array}{rcl}
   H(\xi)&=&\int d^3x \left( \pi^{ab} (\mathcal L_\xi g)_{ab}+ \pi^A \mathcal (L_\xi \phi)_A \right)\\
   S(N)&=&\int d^3x\left(\frac 1{\sqrt{|g|}}\pi^{ab}G_{abcd}\pi^{cd}-(R-2\Lambda)\sqrt{|g|}+H_{\mbox{\tiny matter}}(g_{ab},\phi_A,\pi^A)\right)N(x)\\
G^\alpha(\lambda_\alpha)&=&\int d^3x G^\alpha(g_{ab},\phi_A)\lambda_\alpha\sqrt g
 \end{array}
\end{equation}
where we denote the non-gravitational configuration degrees of freedom collectively by $\phi_A$ and their canonically conjugate momenta $\pi^A$. We assume that the matter Hamiltonian $H_{\mbox{\tiny matter}}$ does neither contain $\pi^{ab}$ nor any spatial derivatives of $g_{ab}$, which seems to be an assumption that is realized in nature. We furthermore assume the constraint associated to internal gauge symmetries to be a functional of only the ``position" variables $\phi^A, g_{ab}$. This is a condition realized in all matter fields studied, and which greatly simplifies treatment. 

To construct a linking theory we extend phase space by a scalar $\phi(x)$ with canonically conjugate momentum density $\pi_\phi(x)$ and impose the additional first class constraint
 $Q(x)=  \pi_\phi(x)$,
which eliminates the phase space extension on shell. To perform canonical best matching for volume preserving conformal transformations, we specify the conformal weight of matter fields by neutral coupling. We thus use the generator
\begin{equation}
 F=\int d^3x\left(e^{4\hat \phi} g_{ab}\Pi^{ab}+\phi \Pi_\phi+\phi_A \Pi^A\right)
\end{equation}
to implement the conformal Kretschmannization. This acts non-trivially only on the gravitational variables. The resulting canonical transformations have $Q(\rho)$ and (respectively) $H(\xi)$ transform weakly into
\begin{equation}
 \begin{array}{rcl}
   Q(\rho)&=&\int d^3x \rho\left(\pi_\phi-4\left(\pi-\langle\pi\rangle\sqrt{|g|}\right)\right)\\
   H(\xi)&=&\int d^3x \left(\pi^{ab}\mathcal (L_\xi g)_{ab} +\pi_\phi \mathcal L_\xi \phi + \pi^A(\mathcal L_\xi \phi)_A\right).
  \end{array}  
\end{equation}
In the cases of interest that we have studied, the gauge constraint decouples from the conformal transformation, so that $T_\phi G^\alpha\propto G^\alpha\approx 0$.

After imposing the best-matching condition $\pi_\phi(x)=0$ we always get an equation of the form:
\be  \{T_\phi S(N), \pi_\phi(x)\}=T_\phi\left[-\frac{3}{2}S(x)+2( \Delta_{\mbox{\tiny matter}}N(x)-\mean{ \Delta_{\mbox{\tiny matter}}N})\right]
\ee
In the general case one has to compute explicitly the entire bracket $\{ T_\phi H_{\mbox{\tiny matter}}, \pi_\phi\} $ and work out the appropriate form of the $\Delta_{\mbox{\tiny matter}}$ to check its invertibility. Let us here restrict to the simpler case occurring when the matter Hamiltonian does not contain spatial derivatives of the metric tensor nor metric momenta. In this case one gets:
\be\label{equ:def:Delta_gen}\Delta_{\mbox{\tiny matter}}:= \nabla^2-\frac{\pi\langle\pi\rangle}{4\sqrt g}-R+\frac{1}{2\sqrt{|g|}}\left(\frac{\delta H_{\mbox{\tiny matter}}}{\delta g_{ab}}g_{ab}+\frac{3}{2}H_{\mbox{\tiny matter}}\right)\ee
On the constraint surface $T_\phi S=0$ and $\mathcal{Q}=0$, the end result is equivalent to taking
\be \Delta_{\mbox{\tiny matter}}\approx (\nabla^2-\frac{1}{12}\langle\pi\rangle^2)-\frac{\sigma^{ab}\sigma_{ab}}{g} 
+\frac{1}{2\sqrt{g}}\left(\frac{\delta H_{\mbox{\tiny matter}}}{\delta g_{ab}}g_{ab}-\frac{1}{2}H_{\mbox{\tiny matter}}\right)\ee
Thus in this case the criterium for invertibility of the operator rests on:
\be\label{equ:inv_criterium}\frac{1}{2\sqrt{g}}\left( \frac{\delta H_{\mbox{\tiny matter}}}{\delta g_{ab}}g_{ab}-\frac{1}{2}H_{\mbox{\tiny matter}}\right)\leq  \sqrt{g} \frac{1}{12}\langle\pi\rangle^2 +\frac{\sigma^{ab}\sigma_{ab}}{ g} 
\ee
in particular if 
\be\label{equ:bound0} \frac{\delta H_{\mbox{\tiny matter}}}{\delta g_{ab}}g_{ab}-\frac{1}{2}H_{\mbox{\tiny matter}}\leq 0\ee
under the previously assumed conditions  the field can always be included in our model.  A useful form of \eqref{equ:bound0} for the specific case of homogeneous scaling is given by the following: if $H_{\mbox{\tiny matter}}\rightarrow e^{n\phi} H_{\mbox{\tiny matter}} $ scales with power $n$, we have the equivalent condition (assuming $H_{\mbox{\tiny matter}}$ is positive):
\be\label{equ:hom_scal}n-2\leq 0
\ee

 In some cases the inequality \eqref{equ:inv_criterium} cannot be attained, in others it implies a bound on the density of the fields, and yet in others it is always obeyed, since \eqref{equ:bound0} is  valid. Examples of the first kind are obtained from fields with four-dimensional conformal coupling, examples of the second kind are obtained by adding a mass potential term to the Hamiltonian, of the form $\psi^2\sqrt g$, and of the third are Yang-Mills and the massless scalar. 
 However, as soon as this restriction is imposed, one can use the implicit function theorem to complete the construction of SD for these matter models.

\section{Conclusions}\label{sec:conclusions}

As a theory possessing 3-dimensional conformal invariance and unique surfaces of simultaneity, Shape Dynamics seems very counter-intuitive in the light of the teachings of Relativity. Furthermore, the existence of fields with mass would seem to undermine conformal invariance. In this paper we have shown that these prejudices are not justified. Applying ``neutral conformal coupling" to the fields - so that they only feel space stretching but have otherwise no independent scaling - our method preserves the fields' canonical constraints for all cases of interest. Furthermore, for all such cases the matter Hamiltonian contains neither metric momenta nor metric spatial derivatives, enabling us to apply our bound \eqref{equ:inv_criterium}. In such cases, the bound gives a straightforward criterium telling us whether a given field  can or cannot  be encompassed by Shape Dynamics. All the constituent fields automatically obey the bound except for the Higgs potential and the cosmological constant, whose value must be bounded. We point out the important fact that the deSitter solution does not obey the bound \eqref{equ:inv_criterium}, hence  it is not a solution for SD  \footnote{We thank Sean Gryb for pointing this out to us. The statement that the model does not obey the bound implies that the dual theory does not gauge fix the refoliations, and we are left with more than one ``global Hamiltonians" in the dual theory. So one could wonder if this ambiguity is in fact not due to the high symmetry content of deSitter space. To this effect we note that we could always make the $\sigma^{ab}\sigma_{ab}$ term in \eqref{equ:inv_criterium} small enough so that no symmetry is present and yet it still breaks the bound. }, and we thus have a first phenomenologically verifiable deviation from standard GR. This work makes the construction of a phenomenologically viable cosmological model imperative for the development of Shape Dynamics.

\section*{Acknowledgements}

 This work was supported in part by the U.S.
Department of Energy under grant DE-FG02-91ER40674.

\end{document}